\documentstyle[11pt,conf_iap,epsf]{article}
\begin{document}


\def \deg    {$^{\circ}$}
\def \etal   {{et al.\thinspace}}
\def \eg     {{e.g.,}}
\def \cf     {{cf.}}
\def \ie     {{i.e.,}}
\def \hub    {$H_{\hbox{\eightrm 0}}$}
\def \hunits {km s$^{\hbox{\eightrm --1}}$ Mpc$^{\hbox{\eightrm --1}}$}
\def \kms    {{\rm km s$^{\hbox{\eightrm --1}}$}}
\def \sec    {$^{s}$}
\def \arcsec {$^{''}$}
\def \arcmin {$^{'}$}
\def \arcsecpoint {$.^{''}$}
\def \tightenlines {\def\baselinestretch{1}\baselineskip 10pt\small}
\let\tighten\tightenlines
\def\gtorder	{\mathrel{\raise.3ex\hbox{$>$}\mkern-14mu\lower0.6ex\hbox{$\sim$}}}
\def\ltorder	{\mathrel{\raise.3ex\hbox{$<$}\mkern-14mu\lower0.6ex\hbox{$\sim$}}}
\def\sb		{{\rm mag~arcsec$^{-2}$}}
\def\area	{${\rm deg}^2$}
\def\kpc	{\hbox{\rm kpc }}
\def\pc		{\hbox{ pc }}
\def\yr		{ \, {\rm yr}}
\def\peryr	{ \, {\rm yr^{-1} }}
\def\vlos	{ v_{\rm los} }
\def\lsim	{ \rlap{\lower .5ex \hbox{$\sim$} }{\raise .4ex \hbox{$<$} } }
\def\gsim	{ \rlap{\lower .5ex \hbox{$\sim$} }{\raise .4ex \hbox{$>$} } }
\def\solar	{ {\odot} }
\def\lsolar	{ {\rm L_{\odot}} }
\def\msolar	{ \rm {M_{\odot}} }
\def\mearth	{ \rm {M_{\oplus}} }
\def\HI		{{H{\sc I}}}
\def\etal	{{\it et~al.}}


%
\heading{THE PLANET COLLABORATION: \\
         Probing Lensing Anomalies} 

\photo{ }

\author{
M. ALBROW$^{1,2}$,  
J.-P. BEAULIEU$^{3}$
P. BIRCH$^{4}$, 
J. A. R. CALDWELL$^{1}$, 
J. GREENHILL$^{5}$,\\
K. HILL$^{5}$, 
S. KANE$^{5,6}$
R. MARTIN$^{4}$, 
J. MENZIES$^{1}$, 
R. M. NABER$^{3}$
J.-W. PEL$^{3}$, 
K. POLLARD$^{1}$,\\
P. D. SACKETT$^{3}$,
K. C. SAHU$^{6}$, 
P. VREESWIJK$^{3}$,
R. WATSON$^{5}$, 
A. WILLIAMS$^{4}$, 
M. ZWAAN$^{3}$\\
(The PLANET Collaboration)
}
{
$^{1}$ South African Astronomical Observatory, P.O. Box 9, 
Observatory 7935, South Africa\\
$^{2}$ Univ. of Canterbury, Dept. of Physics \& Astronomy, 
Private Bag 4800, Christchurch, New Zealand\\
$^{3}$ Kapteyn Astronomical Institute, Postbus 800, 
9700 AV Groningen, The Netherlands\\
$^{4}$ Perth Observatory, Walnut Road, Bickley, Perth~~6076, Australia\\
$^{5}$ Univ. of Tasmania, Physics Dept., G.P.O. 252C, 
Hobart, Tasmania~~7001, Australia\\
$^{6}$ Space Telescope Science Institute, 3700 San Martin Drive, 
Baltimore, MD. 21218~~U.S.A.
}

\bigskip

\begin{abstract}{\baselineskip 0.4cm 
The Probing Lensing Anomalies NETwork (PLANET) is a worldwide 
collaboration of astronomers using semi-dedicated European, South African, 
and Australian telescopes to perform continuous, rapid and precise 
multi-band CCD photometric monitoring of 
on-going Galactic microlensing events.  
As well as providing important additional 
information on the nature, distribution and kinematics of Galactic 
microlenses, PLANET photometry is optimized for the detection of 
Jovian-mass planets orbiting several AU from Galactic lenses.  
The final PLANET database is expected to contain hundreds of variable stars 
sampled at hourly time scales with 1-5\% precision.
}
\end{abstract}

\section{The PLANET Collaboration}

The massive observing programs launched in the early 1990s to detect 
the transitory microlensing of background sources in the Magellanic Clouds 
and Galactic bulge by (moving) foreground stars or dark lenses 
in the Milky Way are now bearing fruit 
(DUO \cite{alard96}, 
EROS \cite{aubourg93}; 
MACHO \cite{alcock93}; and 
OGLE \cite{udalski93}).  
More than 100 microlensing events by compact 
objects in the Galaxy have been reported in the literature so far.  
It has proved possible to distinguish the standard microlensing 
light curve, which assumes that a point-source undergoes 
amplification by a point lens with relative rectilinear motion, 
from those of other forms of stellar variability.  
Unfortunately, the Einstein ring crossing time, 
a degenerate combination of the lens mass, 
and the geometry and kinematics of the source-lens-observer system, 
is the only fitting 
parameter of the standard curve that contains information 
about the lensing system.  
In order to learn more about the microlenses, 
information must be gleaned from the ``anomalous'' fine structure 
in the light curves caused by departures from the standard assumptions. 
Reliable detection of most of these anomalies requires more 
frequent and precise monitoring than that performed by 
the current survey teams.

PLANET ({\bf P}robing {\bf L}ensing {\bf A}nomalies {\bf NET}work)  
is a worldwide collaboration of astronomers with access to 
a network of European, South African, and Australian telescopes \cite{albrow96} 
designed to meet the challenge of microlensing monitoring.   
The primary goal of PLANET is the detection and characterization 
of the lensing {\it anomalies\/}  
that are expected in the presence of binary sources, 
binary lenses, blending, parallax effects due to 
the Earth's motion, and finite source size effects. 
In particular, since the presence of a lens with a planetary system 
can create a detectable perturbation lasting a few hours to a few days, 
intensive microlensing monitoring is a powerful method for the 
detection of extrasolar planets at kiloparsec distances.   
Secondary goals include the search for serendipitous microlensing 
events in PLANET monitoring fields, and galactic structure and 
variable star studies.

\section{The 1995 Pilot Season}

In its 1995 pilot campaign, PLANET had dedicated access to 
four southern telescopes in June-July 1995:  
(1) the ESO/Dutch 0.92m on LaSilla, Chile, 
(2) the Bochum 0.6m also on LaSilla, 
(3) the South African Astronomical Observatory (SAAO) 1.0m at Sutherland, 
     South Africa, and 
(4) the Perth Observatory 0.6m at Bickley in Western Australia.  
The longitude coverage of PLANET telescopes not only 
acted as a hedge against bad weather, but also increased the chances of 
detecting short-term anomalies of the sort expected for planetary 
events and caustic crossings, and allowed independent checks 
on any observed anomaly from another site.  
With the addition of the Canopus 1m in Tasmania, 
PLANET has expanded its longitude 
coverage for the 1996 season, which is now in progress.  

\vskip -0.5cm 
\hsize 5.0in
\hglue 1.3cm
\epsfxsize=\hsize\epsffile{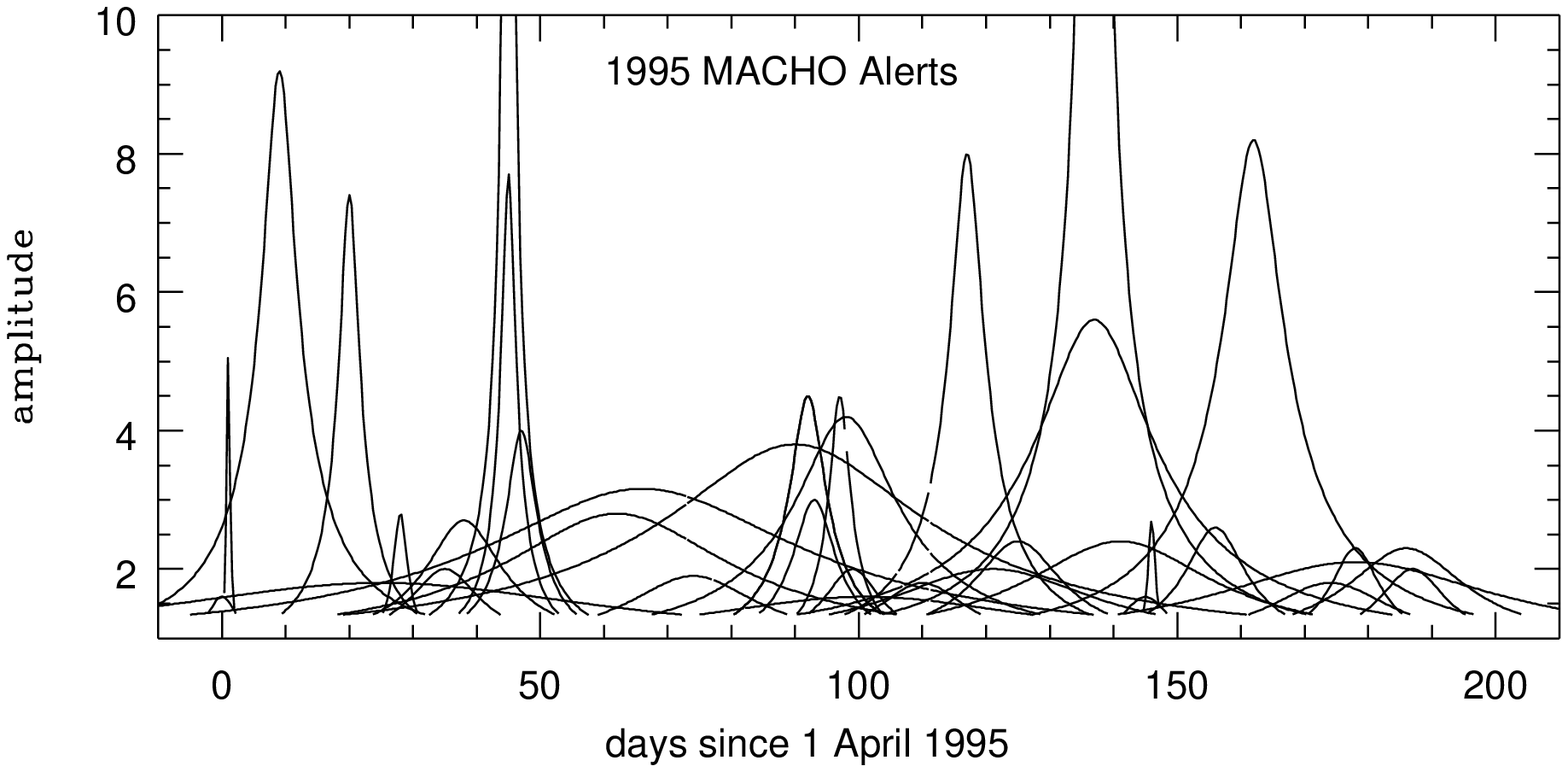}
\vskip -6.5cm
\hglue 0cm
\hsize 17cm

\noindent
{\tighten {\bf Fig.~1:~}  
Light curves for the 1995 real-time electronic alerts given by 
MACHO; parameters are from their alert page at 
http://darkstar.astro.washington.edu.  
The 1995 PLANET pilot season corresponded to days 73-111.  
}

\bigskip

The alert capability of OGLE \cite{udalski94} and MACHO \cite{pratt96}, 
which allows them to detect and electronically report on-going microlensing 
events in real-time, has enabled PLANET to use its small 
fields of view ($\sim$4\arcmin) to best advantage, by 
concentrating on monitoring events already known to be in progress.  
The EROS II collaboration plans to give alerts in the 1997 bulge season.  
The number of 1995 alerts assured that PLANET telescopes were kept 
busy whenever the bulge was visible; the majority of on-going events 
that coincided with the pilot campaign were monitored (Fig.~1).

Microlensing detection teams typically sample a field once a night 
and thus produce light curves that 
are well-sampled on the 10 --- 100 day time scales 
expected for lenses with $0.1 \ltorder M \ltorder 1~\msolar$ with 
photometric precision of about 0.25 mag at V=20 (I=18.5)  
(Cook, these proceedings).  
Analysis of the 1995 pilot data has indicated that 
PLANET telescopes can be continuously and 
fruitfully employed throughout the bulge season performing photometry 
beyond the reach of the microlensing survey teams providing the alerts:   
PLANET photometry is $>$10 times more frequent (Fig.~2) and 
and $\sim$5 times more precise (Fig.~3) than that of the MACHO team, 
for example.  

\medskip

\hsize 4.1in
\hglue 1.85cm
\epsfxsize=\hsize\epsffile{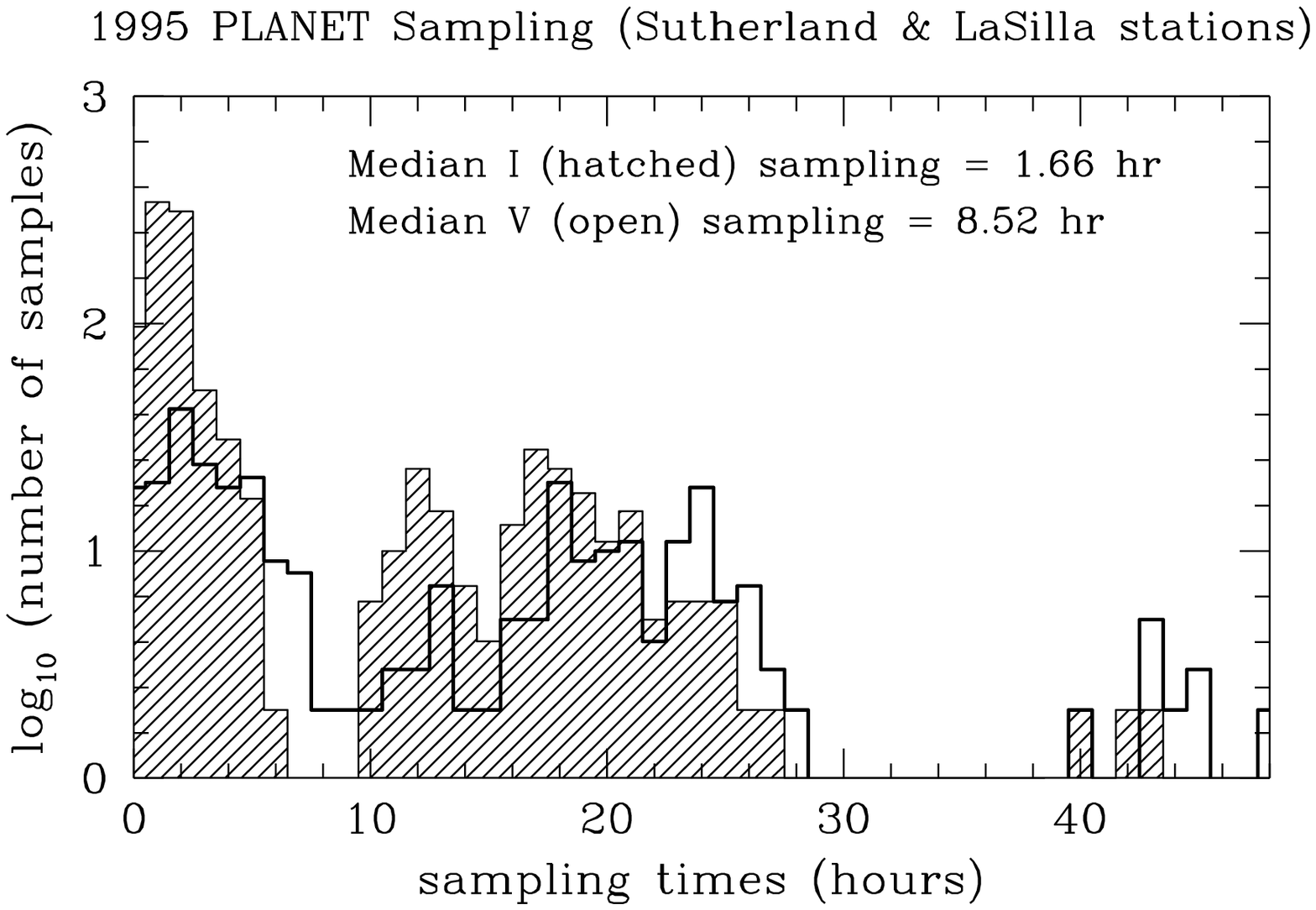}
\vskip -4.45cm
\hglue 0cm
\hsize 17cm

\noindent
{\tighten {\bf Fig.~2:~}
Histogram on a logarithmic scale showing the time between PLANET 
photometric measurements in 1995 for Sutherland (SAAO) and LaSilla 
stations combined in the I (hatched) and V (open) bands.
}

\bigskip

\noindent Relative crowded-field 
photometry was performed to I$\approx$19.5 (V$\approx$21) 
using a set of 10 secondary standards in each field that were 
calibrated during the run.  Exposure times 
were adjusted to obtain comparable errors in I and V.       
Using 1m class telescopes, relative 1\%, 2\% and 6-7\% photometry 
was obtained at I=15, I=17 and I=19 respectively, and was limited by 
crowding, not photon noise.  
The binarity of MB9512, one of the seven events monitored intensively 
in the 1995 campaign, was discovered in real time with PLANET  
mountain-top reduction.~~~Final DoPhot \cite{schechter93} reduction   
yielded the light curve shown in Fig.~4. 
The 1995 data are now fully reduced; a paper is in preparation.

\medskip

\hsize 4.4in
\hglue 1.5cm
\vskip 10.5cm 
\vskip -1.0cm
\hglue 0cm
\hsize 17cm

\noindent
{\tighten {\bf Fig.~3:~} 
Formal DoPhot error as a function of I-band magnitude for one 4\arcmin\ 
PLANET (LaSilla) field taken in 1.5\arcsec\ seeing.  
The number of stars and median error (mags) in each magnitude bin is 
given above that bin. 
{\it Top:~\/} Only stars with well-measured 
point spread functions are shown.  
{\it Bottom:~\/} All measured stars are plotted.}

\hsize 4.5in
\hglue 1.5cm
\epsfxsize=\hsize\epsffile{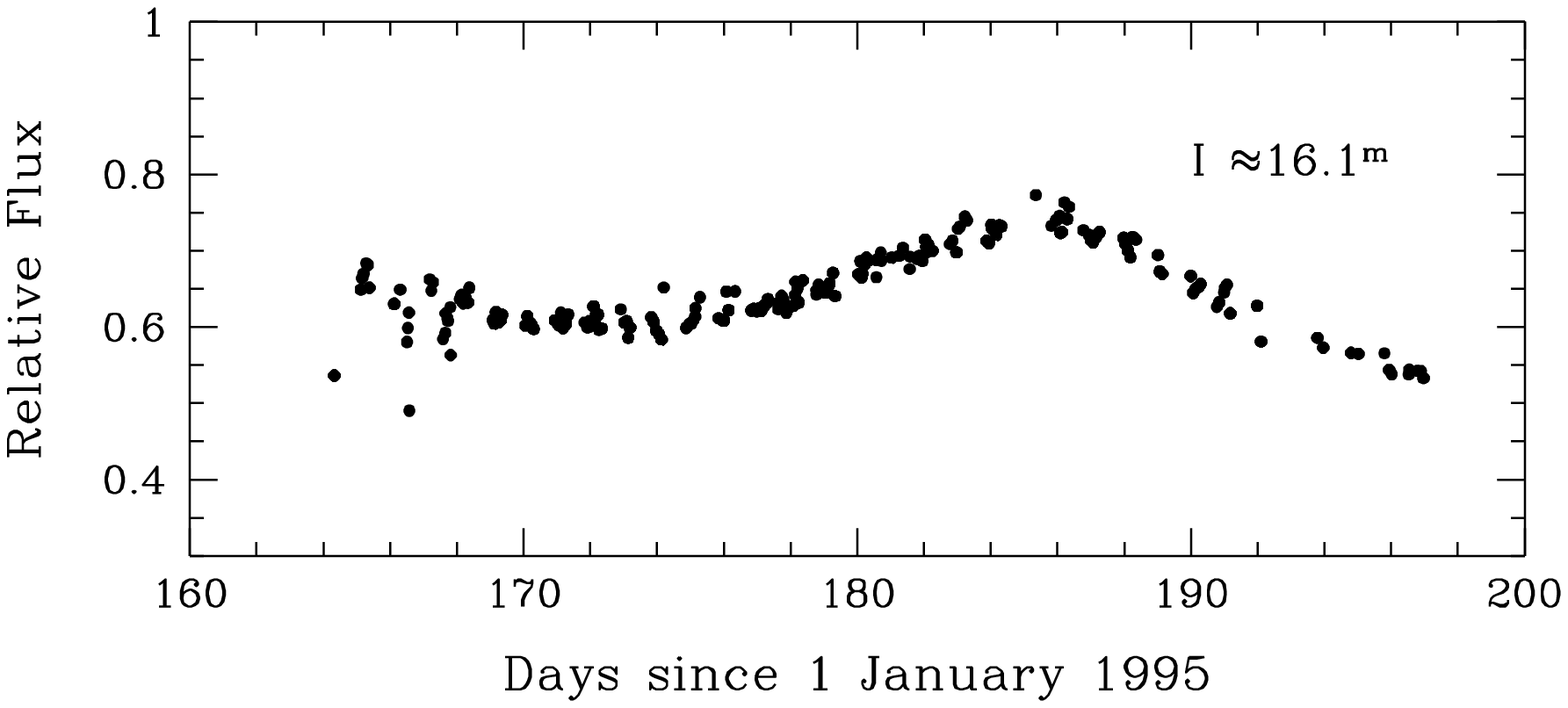}
\vskip -6.3cm
\hglue 0cm
\hsize 17cm

\noindent
{\tighten {\bf Fig.~4:~} 
The relative flux of MB9512 in arbitrary units.  
About 200 {\it unaveraged\/} points from three sites 
are plotted to indicate the true scatter.  
Higher scatter in the 
first nights is due to poor seeing and high background.}

\section{Astrophysical Returns of PLANET Monitoring}

The primary goal of PLANET is the detection of microlensing anomalies 
that will enable the characterization of Galactic microlenses and their 
sources, and the detection and characterization of extra-solar planets. 
Multiple lenses create caustics that result in sharp enhancements 
with durations as short as a few hours.  The resolution of 
these caustic crossings would more tightly {\it constrain the distribution 
of binary masses and separations\/} in the Galactic disk and bulge 
\cite{mao95}.   

A special case of a binary is a lens with 
a planetary system.  A planet in the ``lensing zone'' 
(1-6 AU for most lenses) with a planetary Einstein ring that does not 
resolve the source, will create sharp caustic structure  
in the light curve with durations of a few hours to a few days 
\cite{mao91}.  Photometry capable of characterizing 4-5\% deviations  
would result in detection probabilities near 15-20\% for 
Jupiter-mass planets in the lensing zone \cite{gould92}.  
If extra-solar planetary systems are common, 
precise microlensing monitoring can produce  
{\it distributions of planet-lens mass ratios and 
reduced projected orbital radii for planets 
around randomly-selected stars at kiloparsec distances from Earth.\/} 
On the other hand, non-detection of planetary lensing anomalies 
would place strong constraints on Jupiter- to Neptune-mass planets, 
if $\sim$100 events can be monitored with 
sufficient precision and sampling.
The caustic structure of Earth-mass planets would 
resolve giant sources, resulting in only a fraction of the source 
being amplified and a severe reduction in the size and 
chance of a planetary perturbation \cite{bennett96}.
Reliable detection of Earth-mass planets requires characterization of 
1-2\% deviations against non-giant sources for hundreds of events.

PLANET's photometry is much more suited than that of the 
microlensing surveys to the detection of 
color-shifts due to blended light from chance superpositions \cite{alard96aa}
or from the lens itself \cite{buchalter96apj}.  
Measuring blending 
produces {\it more accurate time scales for microlensing events\/} and 
{\it constrains the mass of stellar lenses\/}.  
The non-rectilinear motion of the Earth around the Sun causes a 
parallax-shift in every event that in principle can be 
used to extract {\it kinematical information about the lens\/}, but 
hourly sampling and $\sim$1\% photometry are required to 
achieve a 10\% detection rate \cite{buchalter96}.  
Lensing of giants at small impact parameter will resolve the source star structure, producing a chromatic signal of 2-4\% at V-I due to 
limb-darkening.  Detection of this small chromatic signal over the 
few hours of the transit would provide {\it a test of model atmospheres 
for giants\/} \cite{gould96apj}.  
In addition, since the physical size of the source 
can be determined, source resolution yields the transit time and 
thus {\it the proper motion of the lens\/}, if the 
light curve is well-sampled over its peak with small photometric 
error \cite{peng97}.

In addition to the characterization of lensing anomalies that 
result from rapid, precise measurements of the microlensing 
event in each PLANET field, a large database of magnitudes 
and colors will be generated for the other 6,000 - 10,000 stars in each 
field.  In particular, the PLANET database is especially well-suited 
to the {\it study of short-period, low-amplitude variables\/} (Fig.~5).  
Variables with sub-day periods have been discovered in the  
pilot season data.   Furthermore, since PLANET monitors 

\hsize 2.9in
\hglue -0.15cm
\epsfxsize=\hsize\epsffile{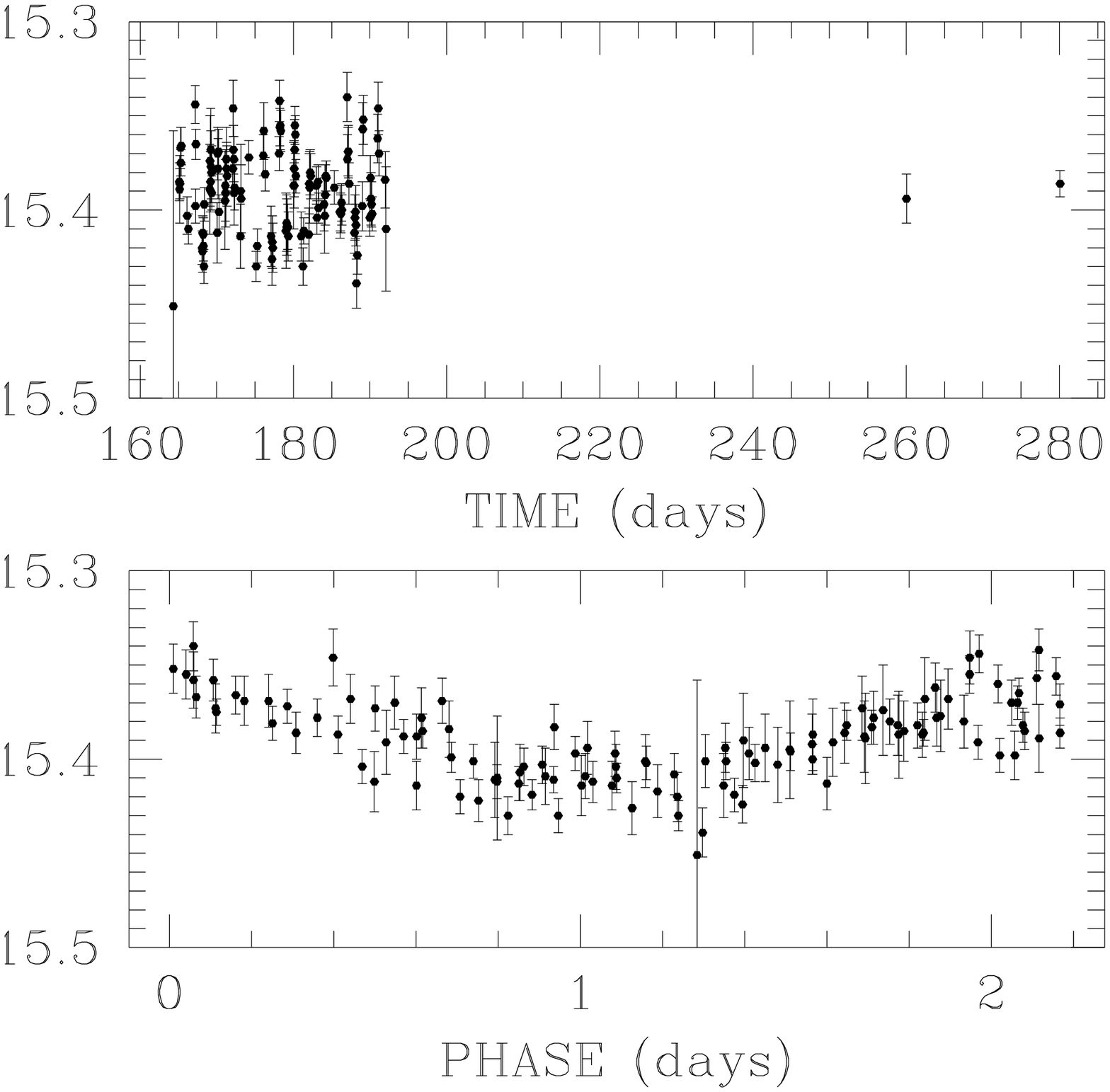}
\vskip -7.4cm

\hsize 2.8in
\hglue 8.4cm
\vskip 6.8cm 
\vskip -0.25cm
\hglue 0cm
\hsize 17cm

\noindent
{\tighten {\bf Fig.~5:~}
{\it Left:~\/} The (LaSilla) I-band 
light curve of a variable found in one of the 
PLANET fields is shown in the top panel and is phase-wrapped in the 
bottom panel to reveal its 2-day period and peak-to-trough amplitude 
of $\sim$0.05 mag. {\it Right:~\/} A PLANET color-magnitude diagram 
containing $\sim$5000 stars for an OGLE-alerted field.    
}

\bigskip

\noindent  in standard 
V and I filters, precise color-magnitude diagrams (CMD) can be built for 
fields scattered throughout the bulge, and the morphology of 
the CMD along different lines of sight utilized to 
disentangle the {\it effects of differential extinction\/}  
from those resulting from {\it Galactic structure\/}.

\vfill
\end{document}